\documentclass{aa}  

\usepackage{graphicx}
\usepackage{lineno} 
\usepackage{lipsum} 
\usepackage{txfonts}
\usepackage[breaklinks=true]{hyperref} 
\hypersetup{
  colorlinks   = true, 
  urlcolor     = red, 
  linkcolor    = blue, 
  citecolor    = blue, 
  breaklinks   = true 
}
\begin{document}

\title{Filtering out large-scale noise for cluster weak-lensing mass estimation}

   \author{C. Murray
          \inst{1,2,3}
          \and
          C. Combet\inst{2}
          \and
          C. Payerne\inst{2,3}
          \and
          M. Ricci\inst{1}
          }

\institute{
Université Paris Cité, CNRS, APC-IN2P3, 75013 Paris, France
\and Université Grenoble Alpes, CNRS, LPSC-IN2P3, 38000 Grenoble, France
\and Université Paris-Saclay, Université Paris Cité, CEA, CNRS, AIM, 91191, Gif-sur-Yvette,
}

   \date{}

 
  \abstract
{We present a new method for estimating galaxy cluster masses using weak-lensing magnification. The effect of weak-lensing magnification introduces a correlation between the position of foreground galaxy clusters and the density of background sources. Therefore, cluster masses can be inferred through observations of these correlations. In this work, we introduce a method that allows us to considerably reduce noise correlations between different radial bins of the cluster magnification signal via a Wiener filtering of our observed magnification field on large scales. This method can reduce the uncertainty on the estimated galaxy cluster mass and it can also be applied to cluster mass estimation for weak-lensing shear. The method was applied to Hyper-Suprime Cam galaxies and CAMIRA clusters detected within the Hyper-Suprime Cam survey (HSC). With HSC data, we find that our filtering method significantly reduces the correlation of noise between radial magnification bins. The estimated cluster mass is consistent between the filtered and unfiltered methods, with similar errors between the two methods as our current measurement errors contain significant contributions from the irreducible shot-noise. For deeper surveys, the effects of shot noise will be less important and this method will lead to greater improvements on the estimated cluster mass. }


   \maketitle
%

\section{Introduction}
\label{sec:introduction}

Galaxy clusters are massive objects and due to their large mass, they produce observable gravitational lensing effects. This can be seen through observations of dramatic Einstein rings and strong lensing arcs, which is referred to as strong gravitational lensing or the deformation of statistical properties of the field of galaxies, referred to as weak gravitational lensing. Weak gravitational lensing can be separated into two distinct effects: weak-lensing shear, where galaxy shapes are sheared and weak-lensing magnification, where solid angles on the sky are magnified. Weak-lensing magnification leads to changes in the observed source size, magnitude, and the apparent number density of sources on the sky. We consider both of these effects in this work, but we focus primarily on weak-lensing magnification.

\begin{figure}
        \includegraphics[width=\columnwidth]{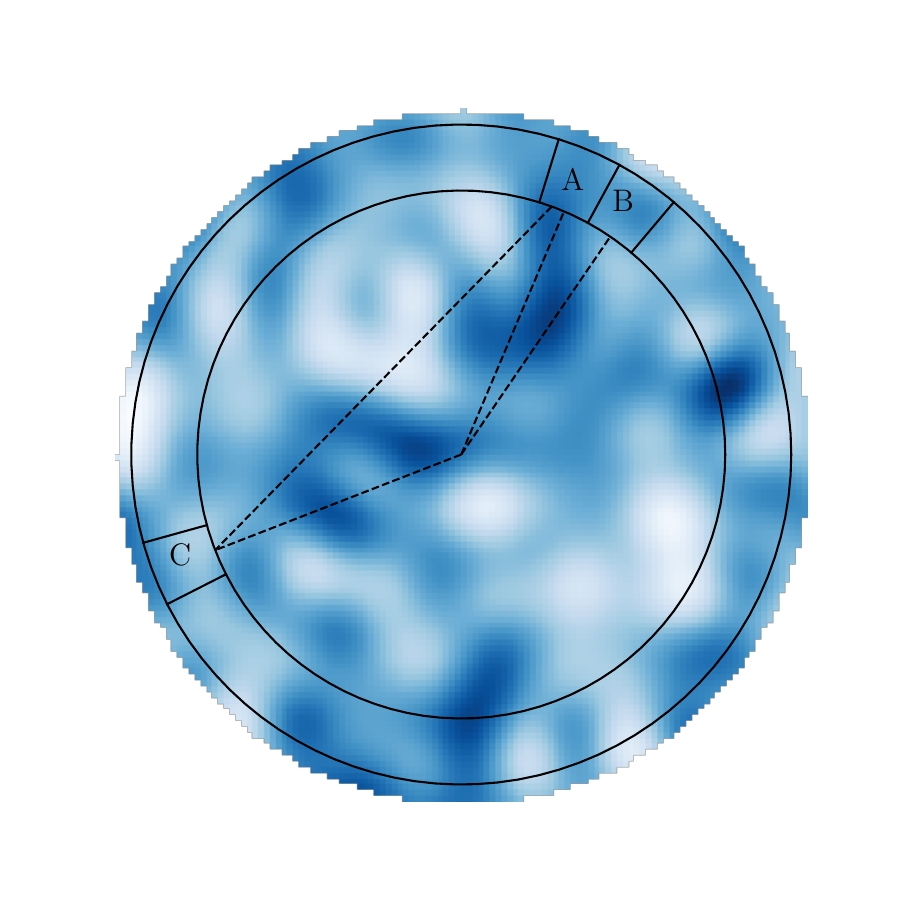}
    \caption{ Illustration of a circular region of a mock galaxy density field, where we consider a circular annulus about a central point. This is a Gaussian noise random field generated from the galaxy angular power spectrum, introduced in Sect. \ref{sec:effective_wiener}.}
    \label{fig:radial_correlation_diagram}
\end{figure}

In general, to estimate cluster masses from the weak-lensing magnification or shear signal, we can take radial averages of noisy measurements. This radial averaging is not optimal when the noise itself has spatial correlations. This is true for both weak-lensing shear and magnification around galaxy clusters provided the measurement error is not dominated by shot noise. Therefore, we should instead try to down-weight or remove the spatial correlations in the shear and magnification fields from the noise. We illustrate an example of the correlations that we wish to minimise in Fig. \ref{fig:radial_correlation_diagram}. Here, we consider a spatially correlated Gaussian random field, which we consider to be our noise field. The correlation is such that nearby regions will be strongly correlated. Therefore, the counts of galaxies in two nearby regions (e.g. the two nearby pixels in Fig. \ref{fig:radial_correlation_diagram} labelled A and B) will be highly correlated, compared to two distant regions (e.g. either of these pixels with the bottom left pixel, labelled C). When taking the radial average, each pixel will contribute equally to the measured signal; however, since these pixels are correlated this average is sub-optimal, they contain the same noise. Beyond this simple illustration, spatial correlations on scales larger than the image will add noise at all scales within the image. For the weak-lensing signal around galaxy clusters, the principle noise contribution (beyond shot noise) can come from scales that are much larger than the typical scales used for weak-lensing analyses of clusters. Importantly, these spatial correlations in the noise introduce a correlation between different radial bins in the magnification signal.

In this work, we develop a Wiener filter approach in order to remove spatial correlations in the galaxy density field in our estimated magnification maps. With this method, we can reduce the impact of uncorrelated structure along the line of sight on the estimation of galaxy cluster masses. This noise, in addition to noise arising from variations in cluster weak-lensing profiles at fixed mass, has previously been referred to as a cosmic variance on cluster mass measurements (\citealp{hoekstra2003well}, \citealp{gruen2015cosmic}); therefore, this approach allows us to move beyond cosmic variance for weak-lensing cluster mass estimations.  Variations in cluster weak-lensing profiles at fixed masses arise from scatter on the mass-concentration relation and clusters are not precisely spherical Navarro-Frenk-White (NFW) profiles (\citealp{nfw}). 

Weak-lensing magnification is a noisier measurement than weak-lensing shear because the intrinsic dispersion of galaxy counts, magnitudes, and sizes used for magnification measurements is greater than the intrinsic dispersion of galaxy shapes needed for shear measurements \citealp{Schneider2000}. Nonetheless magnification provides an important check on weak-lensing shear; as the measurement process is different, we can expect the measurements to be largely independent and provide an important systematic check upon one-another.

Weak-lensing magnification has been used to measure galaxy cluster masses in many studies (e.g. \citealp{taylor1998gravitational,hildebrandt2011lensing,ford2014cluster,chiu2016detection,tudorica2017weak,duncan2016cluster,chiu2020richness}, and \citealp{umetsu2014clash}, who combined shear and magnification). Weak-lensing shear has been used much more extensively to measure galaxy cluster masses (e.g. \citealp{murray2022,2024A&A...689A.252E,Simet2017,Melchior2017,McClintock2019,Bellagamba2019,Linden2014,Applegate2014,Okabe2016} and \citealp{Umetsu2020}).

In this work, we use optical galaxy clusters detected with the CAMIRA algorithm (Cluster finding Algorithm based on Multiband Identification of Red-sequence gAlaxies; \citealt{oguri2018a}, \citealt{oguri2014}) within the Hyper Suprime-Cam (HSC) survey \citep{aihara2019second}. To estimate the weak-lensing magnification signal around CAMIRA clusters, we use galaxies from HSC. Therefore, our dataset is similar to that used in \cite{chiu2020richness}, although we used a more recent release of HSC galaxies for our background sample.

In Sect. \ref{sec:magnification_theory}, we review the effects of weak gravitational lensing around galaxy clusters. Our Wiener filter is introduced for estimating magnification and shear fields with preferable noise properties for cluster mass estimation in Sect. \ref{sec:effective_wiener}. In Sect. \ref{sec:data}, we introduce the HSC data; both the galaxy cluster sample and the background galaxies and their selection used for the magnification measurements. Our measurement methodology is presented in Sect. \ref{sec:measurement_of_magnification_signal}. The stacked magnification masses for both the filtered and unfiltered fields are presented in Sect. \ref{sec:mass_estimation}. We present our conclusions in Sect. \ref{sec:conclusions}.

\section{ Weak lensing by galaxy clusters }
\label{sec:magnification_theory}

Weak-lensing effects are in general not measured on individual galaxies, but are estimated, instead, from the distortion of the properties of a field of galaxies. In this work, we are principally interested in the change of galaxy number counts as a result of magnification and the distortion of galaxy shapes from shear (see \citealt{umetsu2020cluster} for a review on weak-lensing around galaxy clusters).

Weak gravitational lensing distorts the magnitude distribution of galaxies behind clusters, such that the observed distribution, $n_{\rm{obs}}$, can be written in terms of the intrinsic or unlensed distribution, $n_o$, as follows \citep{narayan1989gravitational,Hui2007,Schmidt2009},
\begin{equation}
\label{eq:model}
n_{\rm{obs}}(<m) = n_{\rm{o}}( < m + \delta m ) / \mu ,
\end{equation}
where $n$ refers to the number of galaxies per unit of surface area, $m$ is the galaxy magnitude in a given band, $\mu$ is the weak-lensing magnification, and $\delta m$ is the shift in the magnitude of a galaxy from lensing $\delta m \approx -5 \rm{log}_{10} (\mu) /2$. There are two effects with respect to the unlensed magnitude distribution; a shift from the change in the observed magnitude of galaxies, $\delta m $, and a change in the normalisation, $1/\mu$, from the magnification of the on-sky solid-angle. The magnification can be related to the weak-lensing shear and convergence with
\begin{equation}
\label{eq:magnification}
    \mu = 1/ [ (1-\kappa)^2 - \gamma^2]
,\end{equation}
where $\kappa$ is the convergence and $\gamma$ is the shear. In the limit of low magnification, Eq. \ref{eq:model} can be Taylor-expanded to arrive at a simpler expression,
\begin{equation}
\label{eq:counts_to_magnification}
    n_{\rm{obs}} \approx n_{\rm{o}}(1 +2( \alpha - 1 )\kappa )\,,
\end{equation}
with $\alpha$ the logarithmic slope of the galaxy magnitude distribution at the flux limit of the survey,
\begin{equation}
\label{eq:alpha}
    \alpha = \frac{5}{2} \frac{d \rm{log}_{10} n_{\rm{o}} }{dm}\,.
\end{equation}
The slope $\alpha$ determines how many galaxies are magnified across the magnitude limit of the survey. Therefore, we  see that the number of galaxies can be increased or decreased, depending on whether $\alpha-1$ is positive or negative. The term proportional to $\alpha$ is due to the change in galaxy magnitudes from magnification and the term proportional to $-1$ is due to the reduction of the apparent on-sky density of the galaxies from the magnification of the apparent solid-angle on the sky.

For an azimuthally symmetric gravitational lens, the convergence, $\kappa$, and shear, $\gamma$, can be calculated in terms of the surface mass density (the mass density integrated along the line of sight), $\Sigma$, of the lens,
\begin{equation}
    \kappa(r) = \frac{\Sigma(r)}{\Sigma_{cr}}\,,
\end{equation}
\begin{equation}
    \gamma_+(r) = \frac{\Delta \Sigma(r)}{\Sigma_{cr}}\,,
\end{equation}
where $r$ is the separation from the cluster centre, the subscript $+$ denotes that the tangential shear and we have introduced the excess surface mass density,
\begin{equation}
    \Delta \Sigma(r) = \bar{\Sigma}(<r) - \Sigma(r)
,\end{equation}
where $\bar{\Sigma}(< r)$ is the mean surface mass density within the radius, $r$, and $\Sigma_{\rm cr} $ is the critical surface mass density,
\begin{equation}
\Sigma_{\rm cr} \equiv \frac{c^2}{4 \pi G} \frac{D_s}{D_l D_{ls}}\,.
\end{equation}
Here, $D_s$, $D_l$, and $D_{ls}$ are the angular diameter distances to the source, the lens, and between the lens and source, respectively. Around galaxy clusters, the weak-lensing signal has two principal components; the one-halo term arising from the matter within the cluster and the two-halo term from the correlated large-scale structure around the galaxy cluster. We describe them in turn below.

\subsection{ One-halo term }

To calculate cluster masses from the observed weak-lensing signal, we assume that the cluster mass distribution follows the  NFW profile (\citealt{nfw}). Using this profile, we have analytical equations for the projected surface mass density, $\Sigma(r)$, and the excess surface mass density, $\Delta \Sigma(r),$ in terms of the cluster mass and concentration (\citealt{wright2000gravitational}). Here, $r_s$ is the scale radius of the NFW profile and so, we can define a dimensionless radius $x=r/r_s$, where $r$ is the separation from the cluster centre. We can then express the surface mass density and excess surface density profiles,
\begin{equation}
    \Sigma(x)  = 
    \begin{cases}
      \frac{ 2 r_s \delta_c \rho_c}{ ( x^2-1 )} 
       \left[ 1 - \frac{2}{\sqrt{1-x^2}} \text{ arctanh } \sqrt{ \frac{1-x}{1+x} } \right] 
      \text{ for } x < 1, \\
      
       \frac{ 2 r_s \delta_c \rho_c}{3 } 
       \text{ for } x = 1, \\
       \frac{ 2 r_s \delta_c \rho_c}{( x^2-1 )} 
       \left[ 1 - \frac{2}{\sqrt{x^2-1}} \text{ arctan } \sqrt{ \frac{x-1}{1+x} } \right]  
       \text{ for } x > 1. 
    \end{cases}   
\label{eq:convergence_nfw}
\end{equation}

\begin{equation}
    \Delta \Sigma(x)  = 
    \begin{cases}
       r_s \delta_c \rho_c
       \left[ \frac{8}{x^2\sqrt{1-x^2}} \text{ arctanh } \sqrt{ \frac{1-x}{1+x} } 
        + \frac{4}{x^2} \text{ ln } \frac{x}{2}  
        - \frac{2}{x^2 - 1 } \right.\\ \left.
        + \frac{4}{(x^2-1)\sqrt{1-x^2}} \text{ arctanh } \sqrt{ \frac{1-x}{1+x} } \right] 
      \text{ for } x < 1, \\
      
       r_s \delta_c \rho_c \left[ \frac{10}{3} + 4 \text{ ln } \frac{1}{2} \right]
       \text{ for } x = 1, \\
       
     r_s \delta_c \rho_c
       \left[ \frac{8}{x^2\sqrt{1-x^2}} \text{ arctan } \sqrt{ \frac{x-1}{1+x} } 
        + \frac{4}{x^2} \text{ ln } \frac{x}{2} 
        - \frac{2}{x^2 - 1 } \right. \\ \left.
        + \frac{4}{(x^2-1)\sqrt{1-x^2}} \text{ arctan } \sqrt{ \frac{x-1}{1+x} }  \right]  
       \text{ for } x > 1, 
    \end{cases}   
\label{eq:shear_nfw}
\end{equation}  
where $\delta_c= \frac{\Delta}{3} \frac{c^3}{\rm{ln}(1+c)-c/(1+c)}$ is the characteristic overdensity of the halo, defined by the cluster concentration, $c$, and $\rho_c$ is the critical density of the Universe, $\Delta$ defines the mass and throughout this work, we use $\Delta =200$.

\subsection{ Two-halo term }

At larger angular separations from the cluster centre, the lensing signal from objects clustered around the halo, but not belonging to it will contribute significantly to the lensing signal (\citealt{oguri2011detailed}). This signal generally starts to become significant at approximately 5 Mpc from the cluster centre. 

We can compute the two-halo term following \cite{oguri2011combining} and \cite{oguri2011detailed}, namely,

\begin{equation}
\kappa_{2h} ( \theta , z ) = \int dl \frac{ l }{2 \pi} J_0(l \theta) \frac{\rho_m(z) b_h(z)}{(1+z)^3 \Sigma_{cr} D^2_A(z)} P_m(k_l,z)\,,
\end{equation}

\begin{equation}
\gamma_{+,2h} ( \theta , z ) = \int dl \frac{ l }{2 \pi} J_2(l \theta) \frac{\rho_m(z) b_h(z)}{(1+z)^3 \Sigma_{cr} D^2_A(z)} P_m(k_l,z)\,,
\end{equation}
where $b_h$ is the linear halo bias, $\rho_m$ is the matter density of the Universe, $J_0$ is the zeroth order Bessel function,  $J_2$ is the second order Bessel function and then we have%
\begin{equation}
k_l \equiv \frac{ l }{ (1+z)D_A(z) }\,.
\end{equation}

Here, $\theta$ can be related to the previous expression for the one-halo term expressed as a physical separation using the expression $\theta = r/D_A(z)$. Through the addition of the one and two-halo terms, we can calculate both the shear and magnification signals around galaxy clusters.

\section{ Filtering for cluster mass estimation }
\label{sec:effective_wiener}

In the introduction, we discuss the advantages of taking care when treating spatial correlations of the noise properties of the weak-lensing field. In this section, we describe how we modelled the noise-covariance for our magnification and shear fields and then calculated the variance  this contributes to radial averages of the weak-lensing signal. We then present our Wiener filter-based method to filter out large-scale structure noise as a simplification of a full field-level, maximum, a posteriori estimate of cluster masses. In this context, we discuss other ways in which the cluster mass-richness relation (i.e. an essential part of cluster abundance cosmology) can be estimated at the field level. We estimated the variance in radial annuli from our Wiener filtered fields using mock simulations and we show how filtering large and small scales out of our magnification fields reduces correlations between the radial bins of the magnification signal.

\subsection{ Radial averages and the noise covariance }
\label{subsec:theory_radial_covariance}

Given a correlated noise field described by an angular power spectrum, we can calculate the covariance between different radial bins \citep{schneider1998new,hoekstra2003well,umetsu2011precise}. For example, if we consider the weak-lensing magnification field, $\kappa$. The primary noise contribution, beyond the shot noise, will be the intrinsic clustering of galaxies. Therefore, our noise field is the galaxy angular power spectrum, $C^{gg}_{\ell}/(2(\alpha-1) )^2$, with a multiplicative factor that relates the galaxy count field to the weak-lensing convergence field.  This can be seen by rewriting Eq. \ref{eq:counts_to_magnification} in terms of the galaxy overdensity field, which gives $\delta_{g,\rm{obs}} = \delta_g + 2 \left( \alpha -1 \right) $. Here, $\delta_g = n_o/\bar{n}$ with $\bar{n}$ the mean galaxy density across the survey and $\alpha$ is defined in Eq. \ref{eq:alpha}. 

For the radial average of $\kappa$,
we have
\begin{equation}
\left< \kappa(\theta) \right> \equiv \int d^2 \theta A( |\bar{\theta}|)\kappa(\bar{\theta})
,\end{equation}
with $A$ our radial aperture, $\kappa$ has a covariance between different angular bins $i$ and $j$ given by,
\begin{equation}
C^{\rm{clustering}}_{ij} = \int_{\ell_{\rm{min}}} \frac{\ell d\ell}{2 \pi} C^{gg}_{\ell} \left( \frac{1}{2(\alpha-1)} \right)^2 \hat{J}_0(\ell \theta_i) \hat{J}_0(\ell \theta_j)
\label{eq:noise_covariance}
,\end{equation}
where $\hat{J}_0$ Bessel function is of the first kind and of zeroth order averaged over the annulus,
\begin{equation}
\hat{J}_0(\ell \theta_i) = \frac{2}{(\ell \theta_{i,2})^2-(\ell \theta_{i,1})^2} \left( \ell \theta_{i,2}J_1 ( \ell \theta_{i,2} ) - \ell \theta_{i,1}J_1 ( \ell \theta_{i,1} ) \right).
\end{equation}

For the magnification field, the intrinsic clustering of galaxies will be the largest correlated noise on the field. However there can be other notable contributions, such as the weak-lensing convergence arising from uncorrelated structures along the line of sight or the variance on the profile of the two-halo term contribution. We could also consider noise arising from the survey itself. In practical applications, spatial noise correlations from the survey are also an important noise contribution in HSC (for example \citealt{Nicola2020}).

After estimating the noise contributions of each of these random Gaussian fields to $C^{\left<\kappa\right>}_{ij}$, we can add the shot noise contribution,
\begin{equation}
C^{\rm{shot}}_{ij } = \left( \frac{1}{2(\alpha-1)} \right)^2 \frac{1}{N_{ij} } \delta_{ij}
,\end{equation}
where $N_{ij}$ is the number of galaxies in the radial annulus and, once again, we can convert from galaxy number counts to weak-lensing convergence using Eq. \ref{eq:counts_to_magnification}. The total noise covariance is the sum of these contributions,

\begin{equation}
    C^{n}_{\rm{total}} =  C^{\rm{shot}}_{ij } + C^{\rm{clustering}}_{ij} + C^{\rm{ulss}}_{ij} + C^{\rm{clss}}_{ij} + ...
\end{equation}
Here, $C^{\rm{ulss}}_{ij}$ and $C^{\rm{clss}}_{ij}$ refer to the contribution from uncorrelated and correlated noise, respectively. These contributions can be calculated by replacing the $C^{gg}_{\ell}/(2(\alpha-1) )^2$ term in Eq. \eqref{eq:noise_covariance} with the corresponding angular power spectrum. For the uncorrelated term, this is simply the convergence power spectrum, $C^{\kappa \kappa}_{\ell}$; modelling the angular power spectrum for the two-halo term is somewhat more complicated, but a discussion of this is provided in \cite{gruen2015cosmic}.

The same methodology can be used for the weak-lensing shear. For weak-lensing shear at large-angular separations from the centre, the uncorrelated large scale structure noise contribution will be dominant. At smaller scales the intrinsic variations in the cluster profile become important as well as noise from the surrounding structure (the two-halo contribution). Cluster profiles vary from cluster to cluster at fixed mass and concentration, as clusters are not spherical and perfectly described by NFW profiles.

\begin{figure}
        \includegraphics[width=\columnwidth]{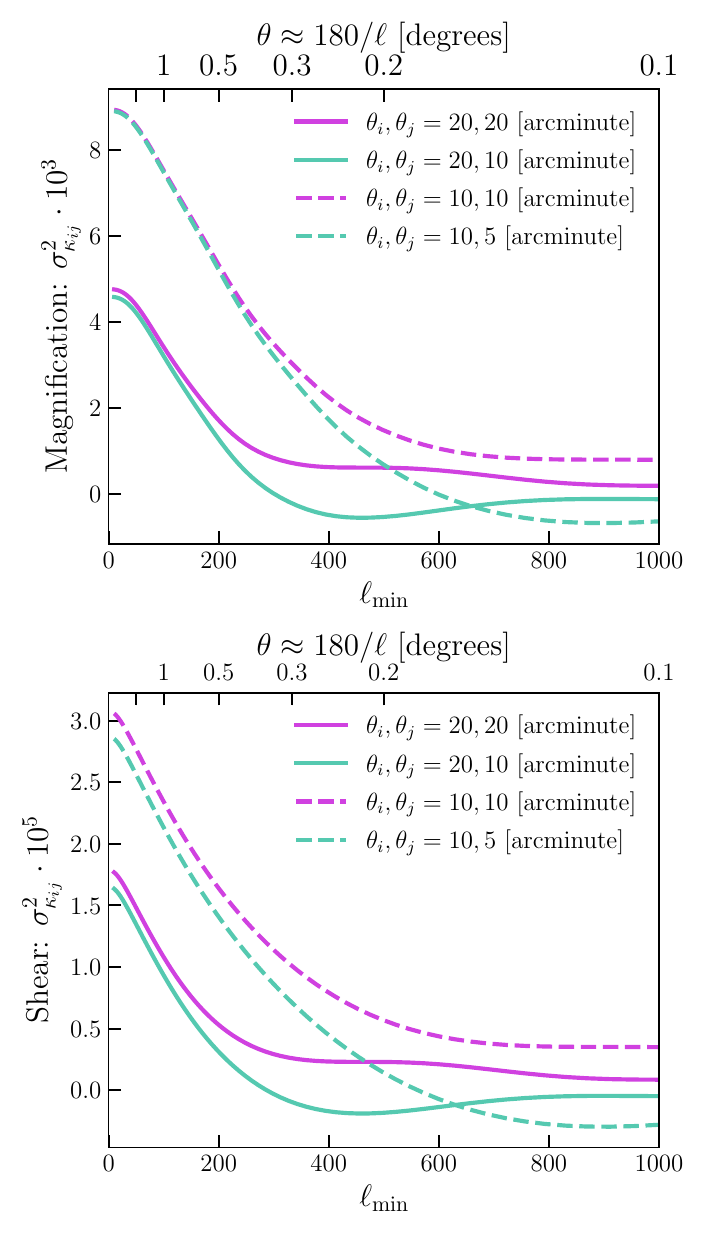}
    \caption{ The noise covariance between radial bins as a function of the minimum $\ell$ used for the integration in Eq. \ref{eq:noise_covariance}. This shows how different multipoles contribute to the radial covariance. $\theta_i$, $\theta_j$ refer to the radial bins used to compute the covariance. The figure shows both the diagonal and off-diagonal terms in the covariance. In the following, we consider the covariance between radial bins of width 1 arcminute. Top: Elements of the covariance matrix between radial bins for the magnification field. Bottom: Elements of the covariance matrix between radial bins for the shear field. }
    \label{fig:noise_scales_of_interest}
\end{figure}

In Fig. \ref{fig:noise_scales_of_interest}, we show results for the covariance between radial averages of the magnification and shear signal calculated using Eq. \eqref{eq:noise_covariance}. We calculate the galaxy angular power spectrum and the shear angular power spectrum as a function of the minimum multipole using the python package \verb|ccl| \citep{Chisari_2019}\footnote{\url{https://github.com/LSSTDESC/CCL}}. We assumed a galaxy bias of $b_g=2$ and a Gaussian probability distribution for the galaxy redshifts, centered on $z=1.4$ with a dispersion $\sigma=0.4/\sqrt{2 \pi}$; for the shear results, we considered a wider redshift distribution of $\sigma=0.5/\sqrt{2 \pi}$. Three points can be observed from this figure; correlated noise from galaxy clustering or the uncorrelated large-scale structure can produce significant correlations between the radial bins, \footnote{This is shown by the off-diagonal covariance terms, for example $\theta_i,\theta_j= 20,10$ having a similar amplitude to the diagonal covariance terms such as $\theta_i,\theta_j= 20,20$.} a significant amount of this noise comes from scales (low $\ell$) much larger than those of interest for cluster lensing (generally smaller than a degree), and the covariance of the magnification field is much larger than that of the shear field. These are important points in allowing us to consider how best to reduce the impact of this noise. 

\subsection{ Field level analysis }
\label{subsec:field_level}

To consider the best way to estimate the magnification and shear fields for cluster mass estimations, a fruitful approach is to pose the question in terms of a maximum a posteriori analysis and then make simplifying approximations.

Therefore, we can consider our data as a pixelised magnification field,

\begin{equation}
    d_i = \kappa_i + n_i
,\end{equation}

\noindent where $i$ refers to the pixel index, $d_i$ is the observed field, $\kappa_i$ is the underlying signal we wish to estimate, and $n_i$ the additive correlated noise. The noise can be approximated to be a Gaussian random variable, however, the magnification signal $\kappa_i$ is not. The noise on the cluster induced magnification field will be primarily composed of shot noise, galaxy clustering noise, and the magnification signal from uncorrelated structures along the line of sight. The galaxy clustering noise on small scales is not Gaussian; however, the noise contributed to the magnification by the galaxy clustering field is integrated over a large redshift kernel along the line of sight and, therefore, the resultant noise field should be close to Gaussian. If we are interested in estimating cluster convergence signal, we can express the Gaussian likelihood as:
\begin{equation}
\label{eq:general_likelihood}
    \mathcal{L}( d_i | \kappa_i ) = \frac{1}{\sqrt{2 \pi |\mathcal{C}|}} \rm{exp} \left[ (d_i - \kappa_i) \mathcal{C}^{-1} (d_i - \kappa_i) \right],
\end{equation}

\noindent where $\mathcal{C}$ is the noise covariance in pixel space.

However, we are not using all the prior information that we have for galaxy clusters. When working with optically detected clusters, we can also include information on the cluster convergence profile as function of mass, $M$, along with the positions and redshifts. Therefore, our likelihood can instead be expressed as:\ 
\begin{equation}
    \mathcal{L}( d_i | M ) = \frac{1}{2 \pi \sqrt{|\mathcal{C}|}} \rm{exp} \left[ (d_i - \kappa_i(M)) \mathcal{C}^{-1}  (d_i - \kappa_i(M) ) \right],
\end{equation}

\noindent where $\kappa_i(M)$ is our model for the convergence field from our known cluster positions and redshifts, namely,

\begin{equation}
\label{eq:mean_mass_kappa}
    \kappa_i = \sum_j \kappa( \theta_i - \theta_j , z_j | M )
,\end{equation}

\noindent where  $j$ is the sum over all of our clusters, $\theta_j$ is the cluster position, and $z_j$ is the cluster redshift. Here, for simplicity, we assume each cluster to have the same mass, but this assumption can be relaxed. For example we could change our model to estimate each galaxy cluster mass and similarly galaxy cluster concentration at the same time. Therefore, our likelihood would change to $\mathcal{L}( d_i | M ) \rightarrow \mathcal{L}( d_i | M_0, M_1,...M_n,  c_0, c_1,...c_n )$ and our model for the convergence would become $\kappa_i = \sum_j \kappa( \theta_i - \theta_j , z_j | M_j, c_j )$. This could be extended to include other parameters of interest, such as the cluster miscentering or ellipticity and orientation. 

Clearly the estimation of so many parameters is non-trivial, not in the least because the magnification field around the individual clusters would only be poorly constrained. Therefore,  we could instead endeavour to estimate directly the mass-richness relation; therefore, our model would become

\begin{equation}
\label{eq:kappa_mass_observable_model}
    \kappa_i = \sum_j \kappa( \theta_i - \theta_j , z_j | \{ \rm{M_0,F_{\lambda},...}\} )
,\end{equation}

\noindent for mass-richness relation parameters $\{ \rm{M_0,F_{\lambda},...} \}$. Here, the covariance matrix would also need to be changed to include the known or estimated scatter around the mean mass-richness relation.

This appears to be the best solution as it allows to estimate the mass-richness relation, which is essential for cosmological analyses for galaxy cluster cosmology at the field level. Importantly, we see that with varying levels of approximation, we can estimate galaxy cluster masses directly on the field level.

\subsection{ Cluster-scale priors }
\label{subsec:filtering}

Instead of the full mechanism of the field-level inference presented in Sect. \ref{subsec:field_level} (which requires precise understanding of the underlying of noise almost definitely beyond the Gaussian approximations we have made), we can simplify our approach by considering a prior on the scales of the galaxy density field of interest to cluster magnification in particular. This allows us to instead reconstruct a cluster magnification map, with preferable noise properties for cluster lensing scales compared to simple galaxy counts. Therefore, we return to the likelihood of Eq. \ref{eq:general_likelihood} and consider a prior on the convergence signal with the knowledge we have of scales relevant for cluster lensing. 

The cluster magnification field is not a Gaussian random variable, however, by placing this as a prior we can create a magnification map with preferable noise properties for cluster scales. Therefore our prior, $P( \kappa_i )$, is that the cluster magnification field follows the probability distribution,

\begin{equation}
\label{eq:kappa_mass_observable_prior}
    P( \kappa_i ) = \frac{1}{\sqrt{|C_{\rm{1h}}|}} \rm{exp} \left[ -\frac{1}{2} \kappa_i C^{-1}_{\rm{1h}} \kappa_i \right],
\end{equation}

\noindent where we determine the scales of interest for the cluster, $C_{\rm{1h}}$, by taking the Fourier transform of the NFW convergence profile given in Eq. \ref{eq:convergence_nfw}. We also set the central region of the NFW, less than 2 arcminutes, equal to $10^{-5}$. This downweights the importance of small scales.

The maximum a posteriori solution (MAP) for this likelihood is the Wiener filter \citep{wiener1949extrapolation}. In Fourier space, we estimated the Wiener filtered field by weighting the observed galaxy density field by the signal and noise covariance as

\begin{equation}
\label{eq:wiener_filter}
    \hat{\kappa}_{\ell m} = \frac{\mathcal{C}^{\kappa}_{1h,\ell}}{\mathcal{C}^{\kappa}_{1h,\ell} + \mathcal{C}^{n}_{\ell}} d_{\ell m}.
\end{equation}

In the following, we demonstrate that the noise properties of this filtered field are preferable to the unfiltered field. Essentially we are using the information presented in Fig. \ref{fig:noise_scales_of_interest}, where low and high multipoles are seen to contribute noise to the cluster magnification field but without contributing to the signal. Therefore by estimating and removing these modes from our magnification maps, we can reduce the noise on our cluster magnification signal. 

\subsection{ Filtered-field covariances }
\label{subsec:forecast}

Using our Wiener filtering approach described in Sect. \ref{subsec:filtering}, we considered  improvements of our Wiener filtered magnification field, as compared to the unfiltered field. We did so by producing simulated magnification fields and shear fields and applying the Wiener filter. The primary aim of this section is to make qualitative, rather than quantitative predictions. We simulated magnification measurements around clusters according to the following steps:\ 
\begin{itemize}
    \item Generate a random Gaussian field given a galaxy angular power spectrum
    \item Draw a Poisson sample for each pixel by assuming a mean number galaxy per pixel
    \item Convert this field to a magnification field with the prefactor $2(\alpha-1),$  as per Eq. \ref{eq:counts_to_magnification}, setting $\alpha=1.5,$ which is a realistic value of $\alpha$ for a galaxy survey
    \item Divide the field into sub-fields, whereby each of these sub-fields are then filtered. This means that the effects of large-scale clustering (beyond the sub-fields) are correctly simulated
    \item Add the cluster magnification signal to the centre of each field
\end{itemize}

Our effective Wiener filter also requires a pixel-noise covariance. We used simulated fields to calculate this as well. We generated a large-field (40 square degrees), subdivided the field into square degree patches, and calculated the pixel-covariance for each sub-field. Then we repeated this multiple times to reduce the measurement uncertainty on the pixel-covariance.

The angular galaxy power spectrum, $C^{gg}(\ell)$, was calculated using \verb|ccl|. As before, we assumed a galaxy bias of $b_g=2$ and a Gaussian probability distribution for the galaxy redshifts, centred on $z=1.4$ with a dispersion $\sigma=0.4/\sqrt{2 \pi}$. This galaxy bias value is similar to the values found in the HSC year 3 analysis \cite{Sugiyama2023} and the redshift distribution is similar to the redshift distribution of sources used for lensing in the HSC year 3 analysis \cite{More2023}. These choices are quantitatively important as they determine the size of the intrinsic fluctuations we aim to reduce with our filtering procedure. The choice of both a large redshift range and a galaxy bias $b_g=2$ will lead to a fairly homogeneous galaxy density field, which reduces the impact of the filtering. Therefore this choice is conservative.

\begin{figure}
        \includegraphics[width=\columnwidth]{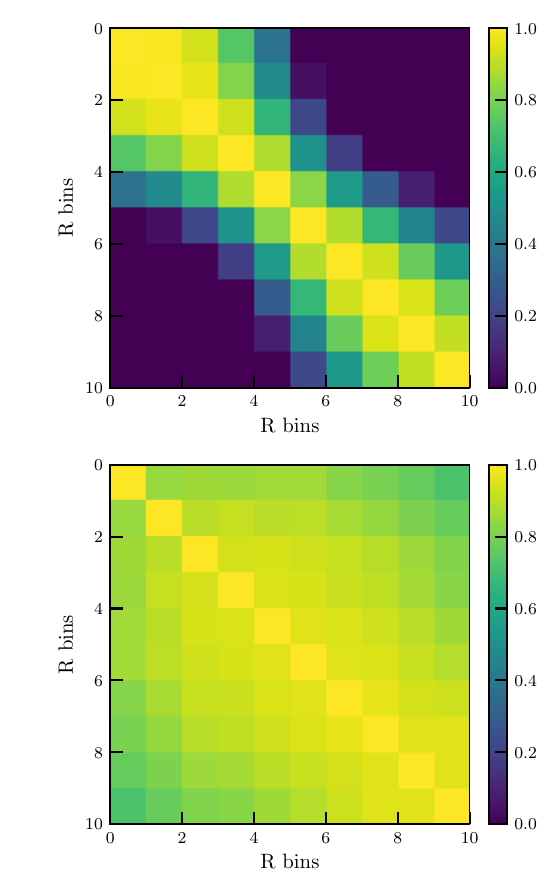}
    \caption{ Top: Radial correlation matrix for the Wiener filtered field. Bottom: Radial correlation matrix for the unfiltered field. For both fields the galaxy density is $N_{\rm{gal}}=20$ per square arcminute. Note: the correlation matrices are limited between $0$ and $1$ (instead of between $-1$ and $1$)  to see   the difference between the two correlation matrices more clearly. Here, the radial bins are linearly spaced bins between 0 and 10 arcminutes.}
    \label{fig:radial_magnification_correlation}
\end{figure}

With this in place we can use the estimated pixel-noise covariance and Wiener filter our fields following Eq. \ref{eq:wiener_filter}.
Using the filtered and unfiltered fields produced from our simulations, we can bootstrap over a set of sub-fields to calculate the radially binned covariance matrix. We show the correlation matrix, namely, the covariance matrix normalised by its diagonal elements, in Fig. \ref{fig:radial_magnification_correlation}. In this test set-up, the filtered fields have significantly smaller off-diagonal correlations. However it is worth noting that the filtering process also changes the magnification signal, therefore improvements in noise properties as measured in the covariance matrix alone do not necessarily demonstrate that the method leads to improvement. For example, the filtering will change the amplitude and shape of the magnification signal. We refrain from presenting information forecasts for each of these cases here, as our simulations are simplified and this would lead to overly optimistic results, compared to real observations.

\section{ Data }
\label{sec:data}

We used galaxy and cluster catalogs obtained from the Hyper Suprime-Cam (HSC) Survey \citep{hsc2022,oguri2018a}. The HSC is a wide-field optical imaging camera operating on the 8.2-m Subaru telescope in Hilo, Hawaii. The survey consists of three depths; wide, deep, and ultra-deep.  In this work we use the wide survey, and only the largest central field of the wide survey. The estimation of the magnification field is easier for a survey with uniform completeness and depth; therefore, the work was simplified by using only the central field. The galaxy sample we use is constructed from the third public data-release of HSC \cite{hsc2022}.

\subsection{ Galaxy cluster sample}

We used the cluster sample from  HSC Cluster finding Algorithm based on
Multi-band Identification of Red-sequence gAlaxies (CAMIRA, \citealt{oguri2018a,oguri2014}).  The cluster finder uses a matched-filter method to find overdensities of red sequence galaxies with a stellar population synthesis model calibrated on spectroscopically observed galaxies. The spectroscopic sample for calibration is constructed from many surveys: SDSS DR12 (\citealt{Alam2015}), DEEP2 DR4 (\citealt{Newman2013}), PRIMUS DR1 (\citealt{Coil2011}), VIPERS PDR1
(\citealt{Garilli2014}), VVDS (\citealt{Fevre2013}), GAMA DR2 (\citealt{Liske2015}), WiggleZ DR1 (\citealt{Drinkwater2010}),
zCOSMOS DR3 (\citealt{lilly2009zcosmos}), UDSz (\citealt{Bradshaw2013}; \citealt{mclure2013sizes}), 3D-HST v4.1.5 (\citealt{Momcheva}), and FMOS-COSMOS v1.0 (\citealt{Silverman}).

 We used rich galaxy clusters, so as to have a pure cluster sample, and only clusters below a redshift of $0.6$, so as to limit cluster member contamination in the background galaxy sample. After applying a richness cut of $N>20$ and a redshift cut of $z<0.6$, we obtained a sample of 1155 clusters within the central region of the survey.  We used the brightest cluster galaxy (BCG) as the cluster centre.

\subsection{ Background galaxy selection }

\begin{figure}
        \includegraphics[width=\columnwidth]{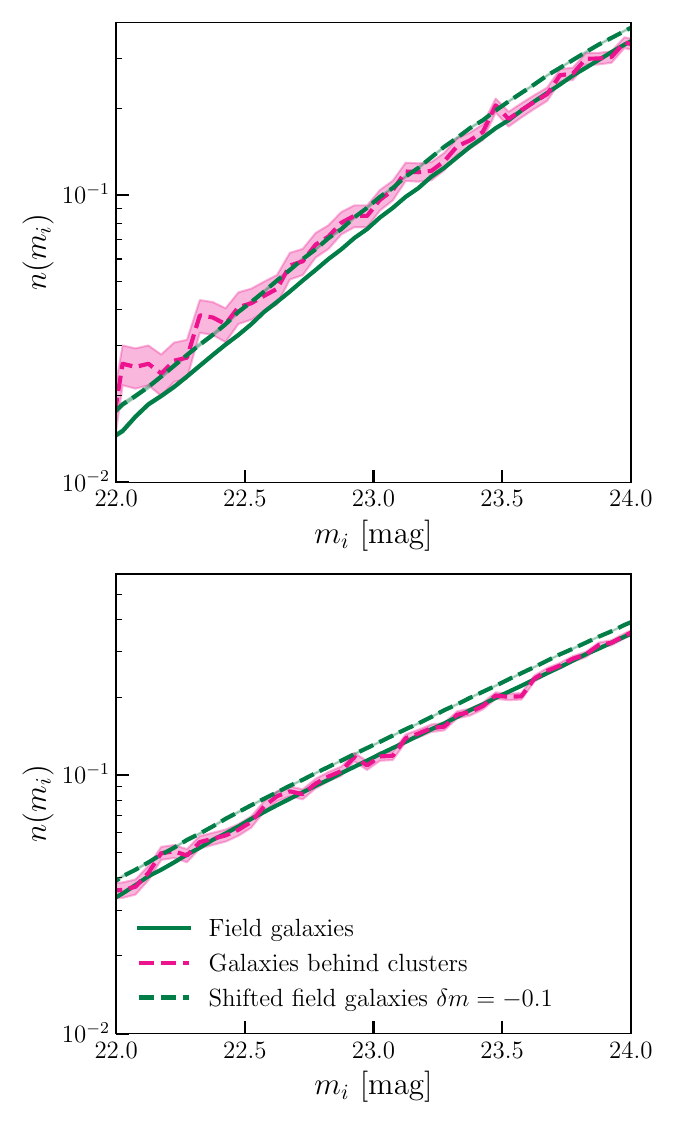}
    \caption{ Normalised magnitude distributions for two different background galaxy sample selections. We compare the magnitude distribution of galaxies in the field as compared to those within the fields of galaxy clusters.  Field galaxies are from random points within the survey footprint. Top: Combined High-z and Low-z galaxy sample selection. Bottom: Photo-z selection. For comparison we add on the figure the shifted field galaxy magnitude distribution. This shift mimics the effect of weak-lensing magnification by the galaxy clusters upon their background galaxies. {The error bars are Poisson noise from the discrete number of galaxies in each magnitude bin.} }
    \label{fig:source_selection}
\end{figure}

For the construction of magnification profiles around galaxy clusters, limiting the contamination of cluster galaxies within the background source catalogue is essential. Additionally, the measurement of photometric redshifts in crowded cluster fields is difficult. We found that selecting background galaxies based on their photometric redshift leads to significant contamination of low-redshift and cluster member galaxies into our background source catalogue. This is shown in Fig. \ref{fig:source_selection} and discussed at the end of this subsection. Therefore, we followed the colour-cut approach of \citealt{chiu2020richness}.

Following \citealt{chiu2020richness}, the galaxy selection was made on the $g-i$ and $r-z$ colour diagram. This allowed us differentiate between foreground, cluster, and background galaxies (\citealt{Medezinski2018}). To limit the shot-noise on our convergence maps, we included as many galaxies as possible. This meant combining both the high and low redshift source galaxies, as defined in \citealt{chiu2020richness}. Therefore, our colour-cuts are the low-redshift source galaxies with cuts,

\begin{equation}
    r-z>0.3,
\end{equation}

\begin{equation}
    1.48 ( r -z ) - 0.71 > g -i,
\end{equation}

\begin{equation}
    -0.625(r-z)+1.06 < g-i,
\end{equation}

and high-redshift source galaxies with cuts,

\begin{equation}
    r-z>0.3
,\end{equation}

\begin{equation}
     \frac{1}{3} ( r -z ) + 0.2 > g -i
,\end{equation}

\begin{equation}
    -0.625(r-z)+1.06 > g-i.
\end{equation}

We also employed a magnitude cut on the $i$-band magnitude with $23.5<i,$ which is essential for reducing the foreground and cluster member contamination. We measured the mean redshift of our background galaxy sample, which is the combination of the low- and high-redshift galaxy samples, by taking the mean of the DEmP\footnote{Direct Empirical Photometric method, \citealt{Hsieh2014}} photo-z mean  for each galaxy from which we obtained $\left< z_{\rm{gal}} \right> = 1.46$. Also, using DEmP, we defined a photo-z selected sample, which comprises all galaxies above a redshift of 1 and below a redshift of 2. The upper limit is set to avoid catastrophic failures of the photo-z often seen in high redshift photo-z estimates.
 
We are able to estimate the effectiveness of our background source selection by looking at the galaxy magnitude distribution before and after the selection of background sources. As explained in Sect. \ref{sec:magnification_theory}, due to the magnification by galaxy clusters, galaxies behind clusters will appear brighter than those in the field. This will shift the galaxy magnitude distribution towards brighter magnitudes. However, the form of the magnitude distribution will not be changed. On the other hand, galaxy cluster member galaxies have a very different magnitude distribution than field galaxies\footnote{Galaxy clusters at low and intermediate redshifts preferentially contain red galaxies.}; therefore, a contamination of the background galaxies by cluster members will distort the magnitude distribution in a manner that is more complicated than a simple shift. In Fig. \ref{fig:source_selection}, for the photo-z selected sample, we see no obvious magnification signal, the magnitude distribution of the galaxies behind the clusters and the field galaxies appear to be the same.  However, for our colour-cut selection, we see the simple shift towards brighter magnitudes in the magnitude distribution, which is to be anticipated by a pure magnification signal. This shows a limitation of the HSC photo-z implementation within crowded cluster fields and for cluster member galaxies. In Fig. \ref{fig:source_selection}, we also see the need for an $i$-band magnitude cut $i<23.5$. Beyond a magnitude of 23.5, the magnitude distribution of galaxies behind the cluster and the field galaxies overlap; therefore, there is no clear magnification signal for these faint galaxies.

\subsection{ Convergence response }
\label{subsec:convergence_response}

The factor of $2(\alpha -1 )$ in Eq. \ref{eq:model} is sometimes referred to as the magnification bias; here, we refer to this as our convergence response. This gives the response of the galaxy density field to the convergence field. We have a magnitude- and colour-limited sample, which can introduce non-trivial magnitude limits on the $r,g, z$ bands which are needed for our colour-cuts. For example, we have a colour-cut $g-i>0.3,$ which (when combined with the $i$-band magnitude cut) introduces a $g$-band magnitude cut of $g<23.8$. As we show in this work, this effect is minimal and the convergence response is primarily determined by the i-band magnitude cut alone.

We can measure  the response on of the magnification directly using the galaxy sample and the selection function without using the necessary approximations inherent in Eq. \ref{eq:model}. Therefore we define a constant, $\mathcal{R}$, which is the magnification response defined as

   \begin{equation}
    \mathcal{R} = \frac{ n_o }{ n_{\rm{obs}} } \frac{1}{\mu} \approx 2 \left( \alpha -1 \right) 
    \label{eq:model_better}
   .\end{equation}

\begin{figure}
        \includegraphics[width=\columnwidth]{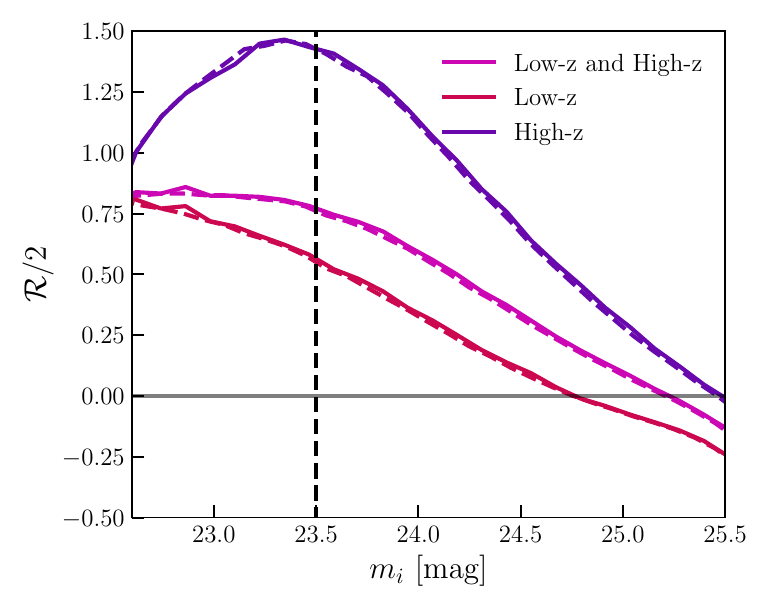}
    \caption{ Response of the galaxy sample to the change in galaxy magnitudes from magnification and dilution. The solid lines are calculated with Eq. \ref{eq:model_better} and the dashed lines are calculated using Eq. \ref{eq:alpha} and the approximation used in Eq. \ref{eq:model_better} . The dashed black line shows the magnitude cut used for our magnification fields, which gives a value of $\mathcal{R}/2=0.77.$ This leads to  $\mathcal{R}=1.54$, which  we used to convert the galaxy overdensity field into a magnification field. }
    \label{fig:alpha}
\end{figure}

Then, $\mathcal{R}$ can be calculated in a simple manner by applying the effects of magnification to our galaxy sample and applying the sample selection function. Therefore, we modulated each of the magnitudes $\{m\}=[m_i,m_g,m_r,m_z]$ with $\delta m = -5/2 \rm{log}_{10} \mu$, with a fiducial $\mu$ and then adjust the normalisation from the magnification of solid angles, with a factor of $1/\mu$ (as explained in Sect. \ref{sec:magnification_theory}). This gives a $\mathcal{R,}$ which does not depend on linear a Taylor expansion (such as Eq. \ref{eq:counts_to_magnification}) and can account for non-trivial sample selection functions, such as the ones included here (noting that a similar approach has been used in \citealt{Hildebrandt2016}). This approach also allows us to take into account that the colours of faint galaxies are not the same as brighter galaxies and the covariance between different photometric-bands is also different. The result of this is shown in Fig. \ref{fig:alpha}, where we also make a comparison to the standard approach, where we directly calculated  the derivative to estimate $\alpha$. The approaches give the same result as in practice the convergence response is driven by the i-band magnitude cut alone. The response demonstrates that our sample is different from that of \citealt{chiu2020richness}, since we used a later data-release of HSC. 

For an i-band magnitude cut at $23.5,$ we obtained a value for our convergence response of $\mathcal{R}=1.54$ for our sample, which is the combination of the high- and low-redshift galaxy samples.

\section{ Measurement of the magnification signal }
\label{sec:measurement_of_magnification_signal}

\subsection{ Convergence map construction }
\label{subsec:magnification_maps}

The galaxy map is constructed by separating the survey into equal area pixels of size 2 arcminutes. We then counted the number of galaxies in each pixel and  calculated our galaxy over-density field,

\begin{equation}
    \delta_g = \frac{n_p}{\bar{n}} - 1
,\end{equation}

\noindent where $\bar{n}$ is the mean galaxy count over all the pixels within the survey mask. Our survey mask was constructed from Arcturus as described in \citealt{Coupon2018}. We used the available random catalogue produced from this mask and applied the same pixel flags as were applied to the galaxy catalogue to the random catalogue. In turn, we created count maps from the random catalogue using the same pixelisation scheme as for the galaxies, namely, using pixels where the random count map is non-zero are regions considered within the survey mask. The count map could be further homogenised, or extended to less uniform regions, by using the deprojection method (e.g. this was applied for galaxy clustering with HSC \citealt{Nicola2020}). Our filtering method itself acts to remove large-scale survey modes, along with large-scale galaxy clustering modes. The density fields were then converted to magnification fields using the convergence value $\mathcal{R}=1.54$ (as given in Sect. \ref{subsec:convergence_response}), such as

\begin{equation}
    \kappa_i = \mathcal{R} \delta_g 
.\end{equation}

Our Wiener filtering was performed on 1 square degree patches around clusters, as opposed to the entire field, which simplifies our estimation of the magnification field noise-power spectrum. We estimated the noise-power spectrum by generating 10,000 random positions within the survey area. For each, we calculated the 2D power spectrum and calculated our noise power spectrum by taking the average of all of these individual estimations. 

We then filtered 1 square degree patches around clusters using a Wiener filter, as explained in Sect. \ref{subsec:filtering}. This provided us with filtered magnification maps around each cluster, which we could then use to estimate the cluster masses. 

\subsection{ Magnification profiles }
\label{subsec:magnification_profiles}

With our magnification maps, for both the filtered and unfiltered fields we created, we measured the stacked magnification profiles around the ensemble of our 1155 CAMIRA galaxy clusters by taking radial averages about the cluster centre. We used $13$ linearly spaced angular bins from $0$ to $30$ arcminutes (although only those between $0.75$ and $5.2$ Mpc are used for the mass estimation).

\begin{figure}
        \includegraphics[width=\columnwidth]{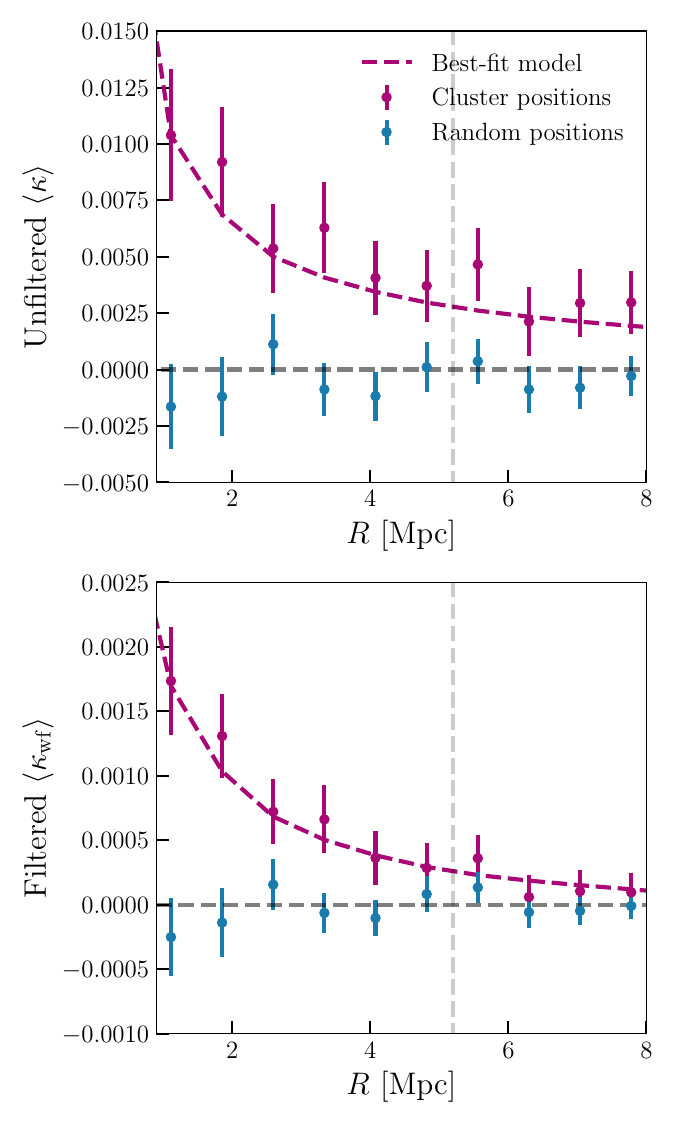}
    \caption{ Top: Convergence profile for the unfiltered convergence field. Bottom: Convergence profile for the filtered convergence field. The error bars on each plot are the standard deviation of the bootstraps over the different cluster profiles. The best-fit model is calculated in Sect. \ref{sec:mass_estimation}. Only points to the left of the gray-vertical line are used for the cluster mass estimation. The mean redshift of the cluster sample is used to convert from the profiles measured in angles (a simpler measurement due to our pixel) to physical units.}
    \label{fig:mean_magnification_profiles}
\end{figure}

We estimated the errors and the covariance matrices for each of our magnification profiles by bootstrapping over each of the different galaxy clusters. We additionally tested our results by performing the same process around 1000 random positions within the survey. For the random positions, we found, for both the filtered and unfiltered profiles, a signal consistent with zero as expected. The magnification profiles are shown in Fig. \ref{fig:mean_magnification_profiles}. We see the same deviations about zero for the mean profiles for the filtered and unfiltered random positions as we used the same random positions for each of these profiles.

\begin{figure}
        \includegraphics[width=\columnwidth]{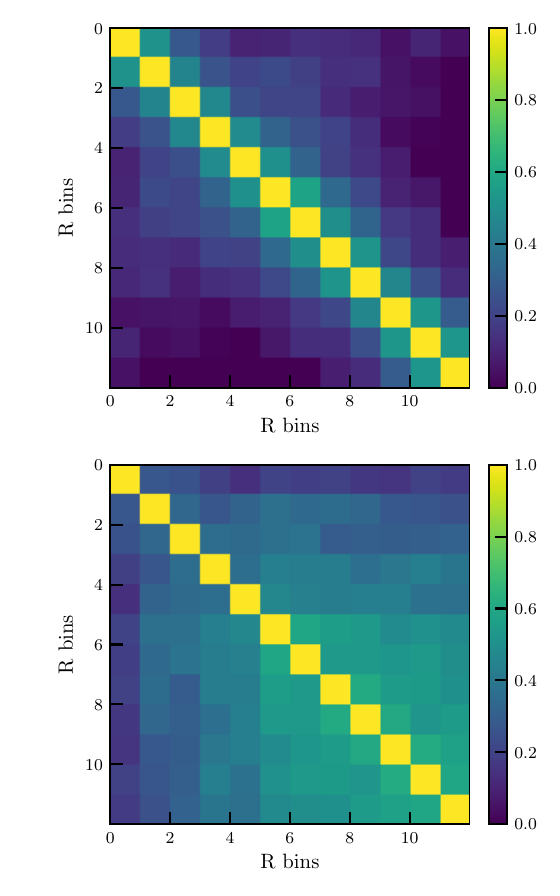}
    \caption{ Top: Wiener filtered field correlation matrix between radial magnification bins. Bottom: Unfiltered field correlation matrix between radial magnification bins. Note: the correlation matrices are limited between $0$ and $1$ (instead of between $-1$ and $1$)  to see the difference between the two correlation matrices more clearly.}
    \label{fig:HSC_radial_magnification_correlation}
\end{figure}

The correlation matrices, computed from bootstrapping on magnification profiles around the galaxy clusters, are shown in Fig. \ref{fig:HSC_radial_magnification_correlation}. We see that the filtered fields significantly reduce correlations between the different angular bins of the cluster magnification profiles.

\section{ Cluster mass estimation }
\label{sec:mass_estimation}

To estimate the stacked galaxy cluster magnification mass, we used a Gaussian likelihood with the covariance and mean magnification profiles as presented in the previous section. For the filtered fields, our model is the sum of NFW magnification and the two-halo term, using the equations presented in Sect. \ref{sec:magnification_theory}. The halo-bias is fixed using the halo-bias relation from \citealt{tinker2010}. Therefore, our likelihood is:

\begin{equation}
    -2 \rm{ln} \mathcal{L}( \rm{log}_{10} M| D ) = \left[ \kappa \left( \rm{log}_{10} M \right) - \hat{\kappa} \right]^T \Sigma^{-1}  \left[\kappa \left( \rm{log}_{10} M \right) - \hat{\kappa} \right]
,\end{equation}
where both the model $\kappa$ and the measurement $\hat{\kappa}$ are changed for the filtered and unfiltered cases. $\rm{log}_{10} M$ is the base-10 logarithm of the lens model mass, $D$ symbolises the data, and $\Sigma$ is the measured covariance matrix, estimated through bootstrapping the magnification profiles. For our mass estimation, we used six linearly spaced angular bins between $0.75$ and $5.2$ Mpc. The inner radius limit was set to avoid the effects of cluster miscentering and higher foreground galaxy contamination in our magnification profiles. The outer radius limit was set to avoid the impact of the two-halo term upon our mass estimation. For the radial range used in this analysis and the constraining power of the data, we were unable to constrain the concentration, halo-bias, and cluster mass concurrently; therefore throughout the analysis, the concentration and halo-bias are fixed. We used the mass-concentration of \citealt{dutton2014} and the halo-bias relation of \citealt{tinker2010} as implemented within the  \emph{Colossus}\footnote{https://bdiemer.bitbucket.io/colossus/index.html} code. 

For the filtered fields, we must account for how the filtering can change the shape of the magnification profile. Therefore, we forward-modelled our magnification signal for the filtered fields. We generated noisy magnification fields using the observed noise power spectrum, following the process used for the mock magnification fields in Sect. \ref{subsec:forecast}. We then Wiener-filtered these fields and measured the mean magnification profile for a given cluster mass for the filtered fields. This forward-modelling approach means that our mass estimation is not biased, provided our noise model describes accurately the observed data.

\begin{figure}
        \includegraphics[width=\columnwidth]{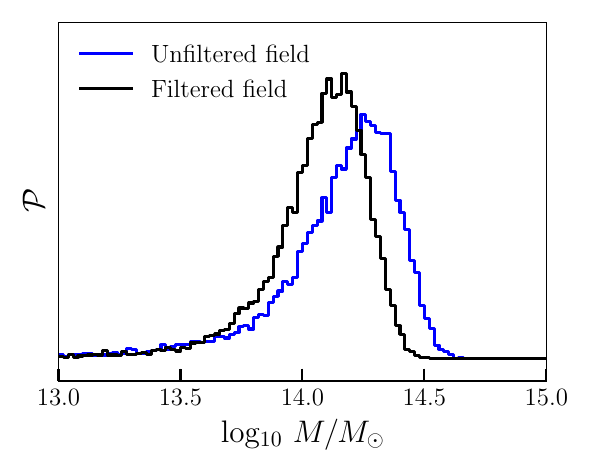}
    \caption{ Posteriors of the mass-estimation for our filtered and unfiltered approach. The posteriors are estimated from the MCMC chains. }
    \label{fig:mass_posteriors}
\end{figure}

Through a Markov chain Monte Carlo (MCMC) analysis using our Gaussian likelihood, we were able to estimate the cluster mass for both the filtered and unfiltered fields. We used the python package EMCEE\footnote{https://emcee.readthedocs.io/}, which is an implementation of an affine-invariant Markov chain Monte Carlo (MCMC) ensemble sampler (\citealt{goodman2010ensemble},\citealt{emcee2013}). For the unfiltered fields, we obtained a mass of log$_{10}$ $M/M_{\odot} = 14.20^{+0.17}_{-0.30}$ and for the filtered fields, we obtained a mass of log$_{10}$ $M/M_{\odot} = 14.10^{+0.15}_{-0.22}$. The posteriors obtained from the MCMC are shown in Fig. \ref{fig:mass_posteriors}. Therefore, given the results for the filtered and unfiltered fields are consistent and give similar errors, we find our mass estimation to be slightly more constraining for the filtered fields.  

\section{Conclusions}
\label{sec:conclusions}

In this work, we  measure stacked galaxy cluster masses using weak-lensing magnification. We present reducible sources of noise on cluster mass estimations, which are relevant for both weak-lensing shear and magnification. With this in mind, we have developed multiple methods, starting from a full field-level analysis, which is able to reduce the uncertainty on the estimated masses of galaxy clusters using weak-lensing information. To proceed, we greatly simplified  this approach through the use of a Wiener filter. Using mock simulations, we were able to show that Wiener filtering offers the possibility to reduce correlations between different radial angular bins of observables averaged in circular annuli. 

This approach was then applied to HSC data, where we were able to slightly reduce the error on the estimation of stacked galaxy cluster masses using the Wiener filter approach, as opposed to the unfiltered field. This slight improvement will become larger with future datasets, as the shot-noise component is reduced as the number of background galaxy sources is increased.

In this work, we have ignored the contribution of dust to the magnification profiles (\citealt{menard2010}). In part because this significantly complicates the work. Ideally, we would expand the work to multi-band photometry, where dust would lead to band-dependent magnification profiles. However, to model the impact of dust, we must also consider the covariance of the galaxy luminosity function and the covariance between the different photometric-bands (\citealt{smith2012}). For example, faint galaxies do not have the same colour as bright galaxies; therefore, as faint galaxies are magnified into the sample, we must be able to account for the different colours of the faint galaxies at the same time as the reddening from dust extinction. 

Whilst this work appears to be the first to use Wiener filtering to explicitly reduce the noise from large-scales in the context of cluster mass estimation, other works have used methods  that lead to similar effects. For example, for the weak-lensing magnification mass estimation in \citealt{umetsuCLASH2014}, the authors used sigma-clipping to reduce noise from angular source-clustering at small scales and normalised the galaxy-density to the mean of the field around the cluster. Also, the mass filters often used in weak-lensing peak studies are perform a local background subtraction, which will also lead to the creation of convergence maps with preferable noise properties, as compared to the unfiltered case (e.g. the truncated isothermal filter developed in \citealt{Schneider1996}  and used in \citealt{oguri2021massfilter} and \citealt{chiu2024}).

This work offers a concrete example of how the noise on galaxy cluster mass estimation can be reduced through the use of more sophisticated methods. We have outlined different ways to estimate the cluster mass-richness relation at the field level. However,  significant work must be done in terms of understanding the noise at small scales in this context, which goes beyond the Gaussian field approximations used within this work.

\begin{acknowledgements}
      Calum Murray acknowledges funding from the French Programme d’investissements d’avenir through the Enigmass Labex.
\end{acknowledgements}

%
%
\bibliographystyle{aa}
\bibliography{aanda} 

\end{document}